\documentclass[10pt,aps,prd, floats, floatfix, superscriptaddress, twocolumn,nofootinbib,amsmath,amssymb]{revtex4-2}

 \usepackage{enumerate}
 \usepackage{graphicx}
 \usepackage{multirow}
 \usepackage{xcolor}
 \usepackage{tablefootnote}
\usepackage{threeparttable}
 \usepackage{bm}
 \usepackage[normalem]{ulem}

 \usepackage{comment}

\usepackage[acronym]{glossaries}
 \usepackage{physics}
 
 {\left\lbrace\begin{array}{@{}l@{}}}%
 {\end{array}\right.}

 \newacronym{frw}{FRW}{Friedmann-Robertson-Walker}
 \newacronym{cmb}{CMB}{Cosmic Microwave Background}

\def\hc{\mathcal{H}}

\begin{document}

\title{Inflation without an Inflaton}

\author{Daniele Bertacca}
\affiliation{Dipartimento di Fisica e Astronomia Galileo Galilei, \\Universit\`a degli Studi di Padova, via Marzolo 8, I-35131, Padova, Italy}
\affiliation{INFN, Sezione di Padova, via Marzolo 8, I-35131, Padova, Italy}
\affiliation{INAF- Osservatorio Astronomico di Padova, \\ Vicolo dell Osservatorio 5, I-35122 Padova, Italy}

\author{Raul Jimenez}
\affiliation{ICC, University of Barcelona, Mart\' i i Franqu\` es, 1, E08028
Barcelona, Spain}
\affiliation{ICREA, Pg. Lluis Companys 23, Barcelona, 08010, Spain.}

\author{Sabino Matarrese}
\affiliation{Dipartimento di Fisica e Astronomia Galileo Galilei, \\Universit\`a degli Studi di Padova, via Marzolo 8, I-35131, Padova, Italy}
\affiliation{INFN, Sezione di Padova, via Marzolo 8, I-35131, Padova, Italy}
\affiliation{INAF- Osservatorio Astronomico di Padova, \\ Vicolo dell Osservatorio 5, I-35122 Padova, Italy}
\affiliation{Gran Sasso Science Institute, Viale F. Crispi 7, I-67100 L'Aquila, Italy}

\author{Angelo Ricciardone}
\affiliation{Dipartimento di Fisica ``Enrico Fermi'', Universit\`a di Pisa, Pisa I-56127, Italy}
\affiliation{INFN sezione di Pisa, Pisa I-56127, Italy}

\begin{abstract}
    
We propose a novel scenario in which scalar perturbations, that seed the large scale structure of the Universe, are generated without relying on a scalar field (the inflaton). In this framework, inflation is driven by a de Sitter space-time (dS), where tensor metric fluctuations (i.e., gravitational waves) naturally arise from quantum vacuum oscillations, and scalar fluctuations are generated via second-order tensor effects. We compute the power spectrum of such scalar fluctuations and show it to be consistent with near scale-invariance. We derive the necessary conditions under which scalar perturbations become significant and much larger than the tensor modes, and we identify a natural mechanism to end inflation via a transition to a radiation-dominated phase. Our proposed mechanism could remove the need for a model-dependent scenario: the choice of a scalar field, as the inflaton, to drive inflation.
\end{abstract}

\maketitle

\section{Introduction}

The inflationary paradigm \cite{Inf0,Inf1,Inf2,MCH,Inf3,Inf4,Linde,Albrecht,Kofman:1985aw} remains the most successful candidate to explain the origin and evolution of the Universe and its large-scale structure\cite{MCH}. This has led to the search among fundamental particle physics for an  inflaton scalar field. So far there has not been any definitive theoretical argument that leads to a single inflaton candidate; in fact, it is possible to build any inflationary potential, even in fundamental theories, like string theory, that fits the observational data: namely the observed tilt of the spectrum of scalar fluctuations, the upper limit on the energy scale of inflation (the yet to be discovered tensor modes)~\cite{Planck18,Galloni:2022mok} and the non-Gaussian  $1-n_s$ signature (e.g., \cite{Bertacca}). This is perhaps the main weakness of the inflation paradigm, as it depends on a model-dependent construction. It is therefore interesting to search for scenarios that are fully model-independent, so that inflation becomes a theory rather than a model. 
Along this line of reasoning, an attempt is done in \cite{GJ1,GJ6,GJ7}, in a framework termed Quantum Fisher Cosmology (QFC). It has been shown that the tilt of the primordial scalar power spectrum can be predicted to be $n_s = 0.9672$ just by considering the Heisenberg uncertainty principle in measuring time in dS. 

Here, we focus on how scalar perturbations are generated in a model-independent fashion, within a purely quantum physics framework. We show that scalar perturbations arise as a second-order effect from tensor perturbations and can become significantly enhanced, allowing them to dominate over the linear tensor modes, which are inherently present in dS. The generation of second-order scalar modes from tensor perturbations was first studied in \cite{Tomita71,Tomita72,Matarrese:1997ay}. Recently, a quantitative analysis of such tensor-induced scalar perturbations was done in \cite{Bari:2021xvf,Bari:2022grh} for post-inflationary epochs. Our scenario relies on a similar mechanism to generate the scalar perturbations. In addition, the instability of dS space \cite{Mottola85,Polyakov,Dvali,Antoniadis_2007} provides both a natural way for a graceful exit from inflation and also the means to end into a radiation dominated epoch. 

If the quantum picture of the dS metric can be described as a Hamiltonian process of scattering and decay of the gravitons composing the coherent state of gravitons, then the self-coupling of gravitons and their coupling with other relativistic particle species leads to metric processes quantum scattering and decay of the constituent gravitons of dS. In this case, the final quantum state cannot be described with coherent states and there will no longer be the dispersion relations of the free quanta propagating on a classical dS background. As pointed out in \cite{Antoniadis_2007}, based on particle creation \cite{Mottola85} and the fluctuation-dissipation theorem \cite{Mottola:1985ee} or, equivalently, on thermodynamic considerations \cite{Mottola:1985qt}, dS spacetime is unstable and the timescale of this instability can be exponentially large given any initial perturbation. This occurs because the geometry of the gravitational field is coupled to the energy-momentum tensor and this quantity, in turn, governs the dynamics of the gravitational field. Consequently, the quantum fluctuations of the vacuum or, equivalently, the effects of particle creation are linked to a background gravitational field.
If we use thermodynamic considerations, see \cite{Mottola:1985qt}, we note that the existence of such a maximally symmetric state cannot guarantee stability against small fluctuations. Indeed, considering a small fluctuation in the Hawking temperature of the horizon—similar to the case of a black hole—leads to a minor net heat exchange between the inner region and both the horizon and its surroundings. 
The entire space is unstable to quantum/thermal fluctuations in its Hawking temperature, nucleating a sort of vacuum bubble at an arbitrary point, breaking global dS invariance \cite{Mottola:1985qt}. See also recent work \cite{Alicki:2023rfv,Alicki:2023tfz}.

The aim of this letter is to introduce this novel mechanism and its main predictions. We derive the exact expressions for the second-order scalar potentials and the scalar power spectrum resulting from second-order tensor perturbations. We demonstrate that the latter agrees with the expected nearly scale-invariance from observations, opening the way for numerous potential follow-up studies and extensions.

\section{Scalar perturbations from tensor modes}

We work on a pure dS metric, which is defined in d-dimensions by the hyperboloid $(dS)_{\alpha} = - X_0^2 + X_1^2 + ... + X_d^2 $ and in Cartesian coordinates, $ds^2 = dt^2 - e^{2t/\alpha} (dx^2 + dy^2 + dz^2)$
where $\alpha \equiv (3/\Lambda)^{1/2}$  and $\Lambda$ is the vacuum energy. We assume Einstein gravity.  

For generality, on the RHS of Einstein's equations, we allow for the presence of a stress-energy tensor which accounts for the sum of the cosmological constant driving our dS expansion plus a generic fluid, with energy density $\rho$, isotropic pressure $p$, four-velocity $u^\mu$ and anisotropic stress tensor $\pi_\nu^\mu$, namely
\begin{equation}
T^\mu_\nu = -{\frac{\Lambda}{8\pi G}} \delta^\mu_\nu + (\rho+p)u^\mu u_\nu+p\delta^\mu_\nu+\pi^\mu_\nu \,,
\end{equation}
As we will see below, the considered fluid unavoidably arises from 
the vacuum expectation value of the second-order contribution to the Einstein's tensor from gravitational waves (GW), which on sub-horizon scales leads to non-vanishing energy, pressure and anisotropic stress. Considering only the tensor contribution we have, in the comoving frame of such a GW fluid,
\begin{equation}
u^\mu=\frac{1}{a}\left(\delta^\mu_0+ \frac{1}{2}v^\mu_2\right)\,,
\end{equation}
from which we can compute $u_\mu$ as:
\begin{align}
&u_0=g_{0\rho}u^\rho=a\left(-1-\frac{1}{2}\psi_2\right);
u_i=\frac{1}{2}a\omega_{2i}+a\left(\frac{1}{2}v_{2i}\right)\,.
\end{align}
For the 00 component at second order we find
\begin{equation}
\frac{1}{2}\delta^2T^0_0=-\frac{1}{2}\delta^2\rho\,.
\end{equation}
 The other components of $T^\mu_\nu$ are given by
\begin{align}
&\frac{1}{2}\delta^2T^i_0=-(1+w)\bar{\rho}\left[\frac{1}{2}v_2^i\right] \,,\\
&\frac{1}{2} \delta^2T^i_j=+\frac{1}{2}\delta^2p\delta^i_j+\frac{1}{a^2}\frac{1}{2}\delta^2\pi^i_j \,,
\end{align}
where $w$ is the equation of state at the background level and the anisotropic stress tensor can be decomposed into a trace-free scalar part, 
$\Pi$, a vector part, $\Pi_i$, and a tensor part, $\Pi_{ij}$, at each
order according to
\begin{equation}
    \pi_{ij}=a^2\left[ \Pi_{,ij}-\frac{1}{3}\nabla^2 \Pi \delta_{ij}+\frac{1}{2}\left(\Pi_{i,j}+\Pi_{j,i}
    \right)+\Pi_{ij}\right]\;.
\end{equation}
\begin{equation}
    \frac{1}{a^2}\delta^2\pi_{ij}= \left[ \Pi_{2,ij}-\frac{1}{3}\nabla^2 \Pi_2 \delta_{ij}+\frac{1}{3}\left(\Pi_{2i,j}+\Pi_{2j,i}
    \right)+\Pi_{2ij}\right]
\end{equation}
we can split the peculiar velocity $v_2^i$ as
$v_{2i}=v_{2,i}+v_{2i}^\perp$, where $v_{2i}^\perp$ is a solenoidal (divergence-free or transverse) vector.

Considering a graviton perturbation, with generic $w$ and $c_s$, where $\delta^2 p=c_s^2\delta^2 \rho$ and manipulating the Einstein equations, we find the potential equation for $\phi_2$, i.e.
\begin{widetext}
\begin{align} 
\label{eq:phi_second_order-all_scales}
\phi_2''& +3\left(1+c_s^2\right)\mathcal{H}\phi_2'
+\left[2\mathcal{H}' +\left(1+3c_s^2\right)\mathcal{H}^2\right]\phi_{2}
-c_s^2\nabla^2\phi_2
 = 8\pi G a^2 \left[ 2\mathcal{H}' +3\left(1+c_s^2\right)\mathcal{H}^2 \right] \Pi_2\nonumber\\
&+ \frac{8\pi G}{3} a^2 \nabla^2 \Pi_2  
+ 8\pi G a^2 \mathcal{H} \Pi_2' 
-\frac{3}{2}\left[2\mathcal{H}' + \left(1+ 3 c_s^2\right)\mathcal{H}^2\right]\nabla^{-4}\partial_i\partial^j {\cal A}^i_j 
- \frac{3}{2}\mathcal{H}\nabla^{-4}\partial_i\partial^j {{\cal A}^i_j}' 
-\frac{1}{2}\nabla^{-2}\partial_i\partial^j{\cal A}^i_j \nonumber \\  
&-\left[2\mathcal{H}' + \left(1+ 3 c_s^2\right)\mathcal{H}^2\right]\nabla^{-2}\Bigg[-\frac{3}{4}\chi_1^{lk,m}\chi_{1kl,m}
- \frac{1}{2} \chi_1^{kl}\nabla^2 \chi_{1lk} 
+ \frac{1}{2}\chi_{1,l}^{km}\chi_{1m,k}^l
+ \frac{1}{2}\chi_{1km}'   {\chi_{1km}'} 
- \frac{3}{2}\nabla^{-2}\partial_i\partial_j \left({\chi_{1k}^{i}}'   {\chi_{1kj}'}\right)\Bigg] \nonumber\\  
&+\frac{3}{8}\chi_{1,k}^{ml}\chi_{1lm}^{,k}
-\frac{3}{8}{\chi_1^{kl}}' {\chi_{1kl}'}
-\frac{1}{4}\chi_{1k,l}^m\chi_{1,m}^{lk} - \mathcal{H}\nabla^{-2}\Bigg[-\frac{3}{2}{\chi_1^{lk,m}}'\chi_{1kl,m} 
+ \frac{1}{2}{\chi_1^{kl}}' \nabla^2 \chi_{1lk} 
- \frac{1}{2}\chi_1^{kl}\nabla^2{\chi_{1lk}}' 
+ \frac{1}{2}{\chi_{1,l}^{km}}'\chi_{1m,k}^l \nonumber\\
&+ \frac{1}{2}\chi_{1,l}^{km}{\chi_{1m,k}^l}'
 -2\mathcal{H}{\chi_{1km}}' {\chi_{1km}'}\Bigg] 
+3\mathcal{H}\nabla^{-4} \left(-2\mathcal{H}{\chi_{1k,j}^{i}}'{\chi_{1k,i}^j}'
+{\chi_{1k,i}^j}' \nabla^2 \chi_{1k,j}^{i}\right) 
+\frac{1}{2}\nabla^{-2}\partial_i\partial_j\left({\chi_{1k}^{i}}'   {\chi_{1kj}'}\right)
\nonumber\\
&+c_s^2\Bigg[-\frac{1}{8}{\chi_1^{ik}}'\chi_{1ki}'
-\hc\chi_1^{ik}\chi_{1ki}'
+\frac{1}{2}\chi^{mk}_1\nabla^2\chi_{1mk}
+\frac{3}{8}\chi^{ml}_{1,k}\chi_{1lm}^{,k}
-\frac{1}{4}\chi_{1k,l}^m\chi_{1,m}^{lk} \Bigg]\;,
\end{align}
\end{widetext}
where
 \begin{align}
{\cal A}^i_j=\frac{1}{2}\chi_1^{lk,i}\chi_{1kl,j}+\chi_1^{kl}\chi_{1lk,j}^{,i} -\chi_1^{kl}\chi_{1l,jk}^i-\chi_1^{kl}\chi_{1lj,k}^{,i} \nonumber \\
+\chi_1^{kl}\chi_{1j,kl}^i 
+\chi_{1,l}^{ik}\chi_{1jk}^{,l}
-\chi_{1,l}^{ki}\chi_{1j,k}^l\;.
\end{align}

Using the traceless part of the $ij$ components we find
\begin{align}
\psi_2=\phi_2-8\pi G a^2 \Pi_2-{{\cal F}_\chi \over 4}\;,
\end{align}
where

\begin{align} 
&{\cal F}_\chi=4 \nabla^{-2}\bigg(
\frac{3}{4}\chi_1^{lk,m}\chi_{1kl,m} + {1\over 2} \chi_1^{kl} \nabla^2 \chi_{1lk}
- {1\over 2}\chi_{1,l}^{km}\chi_{1m,k}^l\bigg)\nonumber\\& -6  \nabla^{-4}\partial_i\partial^j {\cal A}^i_j 
\label{eq:fx}\,,
\end{align}

and, considering only the tensor contribution, the gauge-invariant curvature perturbation on uniform density hyper-surfaces is
\begin{equation}
    \zeta_2=-\phi_2-{\hc \over \bar \rho'}\delta^{2} \rho\,.
\end{equation}
Looking for the particular solution in which $\phi_{2}$ is constant, i.e. on very large scales, and using the energy-momentum constraint equation we get
\begin{align} \label{Psol}
&{8\pi G  \over H_\Lambda^2} \Pi_2 = \eta_{\rm in}^2  \left[ {8\pi G  \over \eta_{\rm in}^2 H_\Lambda^2} \Pi_{2{\rm in}}  
- \left(\phi_{2}-{1 \over 4 } {\cal F}_\chi \right)   \right]\left({\eta \over \eta_{\rm in}}\right)^{\left(5+3c_s^2\right)}
\nonumber \\ 
&+ \left(\phi_{2}-{1 \over 4 } {\cal F}_\chi \right)   \eta^2 \\
 \psi_{2} &=\left[- {8\pi G  \over \eta_{\rm in}^2 H_\Lambda^2}\Pi_{2{\rm in}} + \left(\phi_{2}-{1 \over 4 } {\cal F}_\chi \right) \right] \left({\eta \over \eta_{\rm in}}\right)^{3\left(1+c_s^2\right)}\;.
\end{align}
We can quickly verify that the previous two equations can be combined again to recover the usual relation
\begin{align}
{8\pi G  \over \eta_{\rm in}^2  H_\Lambda^2} \Pi_2 +  \psi_{2} \left({\eta \over \eta_{\rm in}}\right)^2= \left(\phi_{2}-{1 \over 4 } {\cal F}_\chi \right)   \left({\eta \over \eta_{\rm in}}\right)^2\;.
\end{align}
Note that if 
\begin{align}
\left(\phi_{2}-{1 \over 4 } {\cal F}_\chi \right)= {8\pi G  \over \eta_{\rm in}^2 H_\Lambda^2}\Pi_{2{\rm in}}\;,
\end{align}
we will always have  $\psi_{2} =0$ and $\phi_{2}= {\cal F}_\chi/4$; this could also be a solution in our scenario.
Here, $\eta_{\rm in}$ is a suitable initial condition.
Precisely, $\eta_{\rm in}$ is that initial moment in which the hypothesis that $\phi_2$ is constant over time is still correct. Now, considering just a mode with a given comoving $k$, we know that this assumption is reasonably correct just after horizon crossing. $\eta_{\rm in}$ must depend on comoving $k$  (i.e., for different $k$ we should have a suitable $\eta_{\rm in}$). In this case, as a first approximation, we can set $|\eta_{\rm in}(k)| \simeq 1/c_sk$.
If we Fourier transform all these variables we obtain 
\begin{align}\label{psi_2-Fourier}
\tilde \psi_{2} &\simeq \left[- {8\pi G  \over  H_\Lambda^2}{c_s^2 k^2}\tilde\Pi_{2{\rm in}} + \left(\tilde\phi_{2}-{1 \over 4 } \tilde{\cal F}_\chi \right) \right] \left(c_s k|\eta|\right)^{3\left(1+c_s^2\right)}\;.
\end{align}
Now, at superhorizon scales, still assuming $\phi_2$ constant, $\zeta_2$ becomes
\begin{align}\label{zeta-LS}
    \zeta_2=-\phi_2&+{3\hc^3 \over 4 \pi G a^2\bar\rho'} \psi_2=-\phi_2-{H_\Lambda^2 \over 4 \pi G (1+w)\bar \rho} \psi_2\;,
\end{align}
where we left $w$ generic and, let us stress, that is different from $c_s^2$. Including Eq. (\ref{psi_2-Fourier}) and  $$\bar\rho(\eta)= \bar\rho_{\rm in} \left({\eta\over\eta_{\rm in}}\right)^{3(1+w)}=\bar\rho_{\rm in} \left({c_s k|\eta|}\right)^{3(1+w)}\; $$ in Eq. (\ref{zeta-LS}), in Fourier space, it becomes
\begin{align}\label{zeta-sol}
    &\tilde \zeta_2-\tilde\phi_2=-{H_\Lambda^2 \over 4 \pi G (1+w)\bar \rho} \tilde \psi_2 \nonumber \\
   &=-{H_\Lambda^2 \over 4 \pi G (1+w)\bar\rho_{\rm in}} \times \nonumber \\
   &\left[- {8\pi G  \over  H_\Lambda^2}{c_s^2 k^2}\tilde\Pi_{2{\rm in}} + \left(\tilde\phi_{2}-{1 \over 4 } \tilde{\cal F}_\chi \right) \right] \left(c_s k|\eta|\right)^{3\left(c_s^2-w\right)}\;.
\end{align}
Now we need to know the value of  $c_s$ and $\rho_{\rm in}$ at a given mode $k$. It is transparent  that $w-c_s^2>0$ in order for the scalar fluctuations to be larger than the tensor ones. This can be achieved by the same gravitons produced during de Sitter phase \cite{Negro}. For pure radiation with equation of state $1/3$, the scalar perturbations will no longer grow. This provides a natural route to end inflation.

\section{Power spectrum}

As a simple analytical example, let us consider the case where $\psi_{2} =0$, i.e. when $\phi_{2}= {\cal F}_\chi/4$ and we focus only on the first terms in (\ref{eq:fx}); the second term will give a similar dependence. Then the primordial power spectrum of scalar fluctuations $\phi$ is
\begin{equation}
    \langle \phi(\mathbf{k}) \phi(\mathbf{k'}) \rangle = \frac{1}{4}  \langle \phi_2(\mathbf{k}) \phi_2(\mathbf{k'}) \rangle = (2 \pi)^3 \delta^{(3)} (\mathbf{k+k'}) \mathcal{P}_{\phi} (k)
\end{equation}
and 
\begin{align}
  \mathcal{P}_{\phi} (k)  &=  \frac{1}{64 (2 \pi)^3} \frac{1}{k^4} \int {\rm d}^3 k_1 {\rm d}^3 k_2 \;  \delta^{(3)} [\mathbf{k} - (\mathbf{k}_1+{\mathbf{k}_2})] \nonumber \\
&\times \mathcal{K}_h(\mathbf{k}_1, \mathbf{k}_2, k^2)\;
P_h(k_1) P_h(k_2)  \;,
\label{eq:pk}
\end{align}
where the kernel is given by
\begin{widetext}
\begin{align}
\mathcal{K}_h(\mathbf{k}_1, \mathbf{k}_2, k^2)&=
    \Bigg\{ \bigg(k_1^2 +k_2^2 + 3
    \mathbf{k}_1 \cdot \mathbf{k}_2 \bigg)^2 \bigg[\left(1 -  \hat{\mathbf{k}}_1 \cdot \hat{\mathbf{k}}_2 \right)^4 + \left(1 + \hat{\mathbf{k}}_1 \cdot \hat{\mathbf{k}}_2\right)^4 \bigg]  \nonumber \\ 
 &   + 8
 \left(\mathbf{k}_1 \cdot \mathbf{k}_2  \right) \left( k_1^2 + k_2^2 + 3
 \mathbf{k}_1 \cdot \mathbf{k}_2  \right) \left(\hat{\mathbf{k}}_1 \cross \hat{\mathbf{k}}_2\right)^2 
 \left [ 3 + \left(\hat{\mathbf{k}}_1 \cdot \hat{\mathbf{k}}_2 \right)^2   \right ] + 8
 k_1^2  k_2^2  \left(\hat{\mathbf{k}}_1 \cross \hat{\mathbf{k}}_2\right)^4 \left [ 1 + \left(\hat{\mathbf{k}}_1 \cdot \hat{\mathbf{k}}_2 \right)^2 \right ] 
\Bigg\}\;, 
\label{eq:PS}
\end{align}
\end{widetext}
defining $P_h (k)=2\pi^2 \Delta^2_h(k)/k^3$, the tensor power spectrum in dS reads \cite{Watanabe:2006qe}
\begin{equation}
    \Delta^2_h (k) = \frac{16}{\pi} \left (  \frac{H_{\rm inf}}{m_{\rm pl}}   \right )^2\;.
\end{equation}
Here $H_{\rm inf} = \sqrt{\Lambda/3}$ is the Hubble constant during inflation.

Note that the kernel differs distinctly from that of standard single-clock inflation (which for this case is in the bispectrum). This difference will also affect the bispectrum of scalar perturbations, thereby imprinting a unique signature in the squeezed non-Gaussian features of the large-scale structure~\cite{Bertacca}. 
As a consequence, this provides a means to observationally test our scenario. The power spectrum is nearly scale invariant; this can be seen easily as no uncompensated power of $4$ terms of $\mathbf{k}$ appear in the scaling of the momenta in (\ref{eq:PS}).

\section{Discussion}

We have shown that it is possible to generate nearly scale-invariant scalar adiabatic perturbations in pure de Sitter. This is a scenario where the inflaton does not exist, and thus opens up the possibility to provide a picture of inflation that is model independent. There are some interesting features in our calculation. First, note that the scalar fluctuations are generated outside the horizon from the tensor perturbations. Assuming these fluctuations are adiabatic, as in the standard picture of inflation, the perturbations induced in the potential $\phi$ persist after the transition from the dS phase to a radiation dominated epoch through the decay of dS. Similarly to scalar modes, vector perturbations are also generated from second order tensor fluctuations. As expected, the fluctuations deviate from Gaussianity and exhibit a certain intrinsic level of non-Gaussian features, the quantification of which will be addressed in a future publication. 
However, we note the following: 
for any non-conformally invariant field (with mass $m \ll H$) the quantum generation of particles from the vacuum state will add extra contributions to the second-order scalar perturbation mode, hence contributing to a gaussianization by the central limit theorem. 
In our scenario, the magnitude of fluctuations can be determined from eqs \eqref{eq:pk}–\eqref{eq:PS}, which rely exclusively on dS features and potential physical mechanisms that enhance scalar perturbations. These factors suggest a possible explanation for why the fluctuations are at the $10^{-5}$ level.

\begin{acknowledgments}
Funding for the work of RJ was partially provided by
project PID2022-141125NB-I00, and the “Center of Excellence Maria de Maeztu 2020-2023” award to the
ICCUB (CEX2019- 000918-M) funded by MCIN/AEI/10.13039/501100011033.  
DB  and SM acknowledge support from the COSMOS network (www.cosmosnet.it) through the ASI (Italian Space Agency) Grants 2016-24-H.0, 2016-24-H.1-2018 and 2020-9-HH.0.
\end{acknowledgments}

\bibliographystyle{apsrev4-2}

\end{document}